\documentclass{aa}
\usepackage{graphicx,amsfonts,amssymb,natbib}
\begin{document}
\title{Limits on Galactic Dark Matter \\
with 5 Years of EROS SMC Data
\thanks{Based on observations made at the European Southern Observatory,
La Silla, Chile.}}

\author{
C.~Afonso\inst{1,4,8}, 
J.N.~Albert\inst{2},
J.~Andersen\inst{6},
R.~Ansari\inst{2}, 
\'E.~Aubourg\inst{1}, 
P.~Bareyre\inst{1,4}, 
J.P.~Beaulieu\inst{5},
G.~Blanc\inst{1},
X.~Charlot\inst{1},
F.~Couchot\inst{2}, 
C.~Coutures\inst{1}, 
R.~Ferlet\inst{5},
P.~Fouqu\'e\inst{9,10},
J.F.~Glicenstein\inst{1},
B.~Goldman\inst{1,4,8},
A.~Gould\inst{1,7},
D.~Graff\inst{7},
M.~Gros\inst{1}, 
J.~Haissinski\inst{2}, 
C.~Hamadache\inst{1},
J.~de Kat\inst{1}, 
T.~Lasserre\inst{1},
L.~Le Guillou\inst{1},
\'E.~Lesquoy\inst{1,5},
C.~Loup\inst{5},
C.~Magneville \inst{1}, 
J.B.~Marquette\inst{5},
\'E.~Maurice\inst{3}, 
A.~Maury\inst{9},
A.~Milsztajn \inst{1},  
M.~Moniez\inst{2},
N.~Palanque-Delabrouille\inst{1}, 
O.~Perdereau\inst{2},
L.~Pr\'evot\inst{3}, 
Y.R.~Rahal\inst{2},
J.~Rich\inst{1}, 
M.~Spiro\inst{1},
P.~Tisserand\inst{1},
A.~Vidal-Madjar\inst{5},
L.~Vigroux\inst{1},
S.~Zylberajch\inst{1}
}

\institute{
DSM/DAPNIA, CEA/Saclay, 91191 Gif-sur-Yvette Cedex, France
\and
Laboratoire de l'Acc\'{e}l\'{e}rateur Lin\'{e}aire,
IN2P3 CNRS, Universit\'e Paris-Sud, 91405 Orsay Cedex, France
\and
Observatoire de Marseille,
2 pl. Le Verrier, 13248 Marseille Cedex 04, France
\and
Coll\`ege de France, Physique Corpusculaire et Cosmologie, IN2P3 CNRS, 
11 pl. M. Berthelot, 75231 Paris Cedex, France
\and
Institut d'Astrophysique de Paris, INSU CNRS,
98~bis Boulevard Arago, 75014 Paris, France
\and
Astronomical Observatory, Copenhagen University, Juliane Maries Vej 30, 
2100 Copenhagen, Denmark
\and
Departments of Astronomy and Physics, Ohio State University, Columbus, 
OH 43210, U.S.A.
\and
Department of Astronomy, New Mexico State University, Las Cruces, NM 88003-8001, U.S.A.
\and
European Southern Observatory (ESO), Casilla 19001, Santiago 19, Chile
\and
Observatoire de Paris, LESIA, 92195 Meudon Cedex, France
}
\offprints{\\Eric.Aubourg@cea.fr, Nathalie.Palanque-Delabrouille@cea.fr}

\date{Received 00 September 2002 / Accepted 00 September 2002}

\abstract{
Five years of \eros\ data towards the Small Magellanic Cloud have been
searched for gravitational microlensing events, using a new, more
accurate method to assess the impact of stellar blending on the
efficiency.  Four long-duration candidates have been found which, if
they are microlensing events, hint at a non-halo population of lenses. 
Combined with results from other
\eros\ observation programs, this analysis yields strong limits on the amount of
Galactic dark matter made of compact objects. Less than 25\% of a
standard halo can be composed of objects with a mass between
$2\times10^{-7}\,\rm M_\odot$ and 1 $\rm M_\odot$ at the 95\% C.L.

}

\authorrunning{EROS Collaboration}
\titlerunning{Limits on Galactic Dark Matter with 5 Years of EROS SMC Data}

\def\ie{{\em i.e.}}
\def\eros{{\sc eros}}
\def\macho{{\sc macho}}
\def\lmc{{\sc lmc}}
\def\smc{{\sc smc}}
\def\sm98{{\sc sm{\footnotesize 98}}}
\def\ccd{{\sc ccd}}
\def\etal{{et al.}}

\maketitle

\section{Research context}
The idea of using gravitational microlensing as a tool to probe the
dark matter in the Galactic halo goes back to 1986 with a paper by
B. Paczy\'nski \citep{Pac86}. Since then, several experiments have
been monitoring millions of stars towards the Large and Small
Magellanic Clouds (\lmc\ and \smc) and candidates have been observed
towards the two targets \citep{alc93, aub93, ans96, alc97a, smc2,las00}.

The detection of candidates towards the \lmc\ suggested a population
of lenses accounting for $\sim$20\% of a standard halo, with a most
probable mass between 0.2 and 0.9~M$_\odot$~\citep{lmc5macho}. 
Results by \cite{las00}  show that objects up to 1~M$_\odot$ cannot 
account for more than 40\% of a standard  halo.  
These lenses cannot be ordinary Galactic stars, 
whose density is much too low to account for the observed optical depth, and
\cite{gol02} demonstrated they cannot be halo white dwarfs with hydrogen atmosphere.
Much debate has occurred on the nature and the location of the lenses, 
the main two possibilities being 0.4~M$_\odot$ dark objects in the halo or 
stars at the low mass end of the main sequence in the clouds themselves~\citep{sahu94, wu94}.

This makes the \smc\ a very valuable target. While the characteristics
(optical depth and duration) of events towards the two clouds should
be similar for halo lenses~\citep{sacgou93}, the different dynamical
properties of the clouds could account for differences if the events
are due to self-lensing.

The \eros\ 2 experiment has been surveying the \smc\ since 1996. We
present here an analysis of five years of data accumulated towards
this target.

\section{Experimental setup and SMC observations}

The telescope, camera, telescope operations and data reduction are as
described in \citet{smc1}, hereafter \sm98, and references
therein. Since July 1996, 10 one-square-degree fields have been
monitored towards the \smc. The first 5 years of data from 8.6 square
degrees spread over these 10 fields have been analyzed. The data set contains
about 5.2 million pairs of light curves in two wide pass bands called thereafter ``red'' and ``blue'', sampled on average once every 2.5 days
from end-April to mid-March when the \smc\ is visible. About 400--500
images of each field were taken, with exposure times ranging from 5
minutes in the center of the \smc\ to 15 minutes in the outermost
region.

\section{Data analysis}\label{sec:analysis}

The analysis of the five-year data set is similar to those of the
first year (\sm98) and of the first two years \citep{smc2} of data
 where details can be found. 

A first set of cuts is designed to select light curves that exhibit a
single significant fluctuation.  This already excludes 99.8\% of the
stars, most of which exhibit flat light curves.

A second set of cuts rejects stars that are a priori likely to be variable, based upon their
position in a color-magnitude diagram: bright blue stars in the upper
main sequence with low-amplitude variations and stars much brighter and
redder than those of the red clump are rejected. These stars are
contained in two thinly populated regions of the color-magnitude
diagram, so rejecting them does not reduce much the number of light
curves on which we can search for microlensing. Stars exhibiting a
strong correlation between the red and the blue light curves outside
the period containing the main fluctuation are also
rejected. Supernovae are rejected by the comparison of the rising and
falling times of the variation.  Half of the light curves
surviving the first set of cuts are rejected at this stage, leaving
about 5000 light curves.

A third set of cuts improves the signal-to-noise ratio of the set of
selected candidates by comparing the measurements with the best-fit
point-source point-lens microlensing light curve. These criteria are
sufficiently loose not to reject light curves affected by blending,
parallax and most cases of binary lenses or sources. The remaining
sample consists of about 30 light curves exhibiting a clean and unique
fluctuation with a smooth time variation.

A fourth set of cuts constrains the time coverage of the event,
excluding in particular all events with Einstein radius crossing times
$t_{\rm E}$ exceeding 1200~days.

The final criterion is as follows: if blending significantly
improves the fit, then the fitted blending should be physical, \ie\
indeed correspond to the amplification of only a fraction of the
total flux recovered. 

To limit contamination of the set of selected candidates by spurious
events, the levels of the above cuts are set such that most variable 
stars are rejected by at least two of them. The tuning of each cut 
and the estimation of the efficiency of the analysis is done with Monte 
Carlo simulated light curves, as described in \sm98. Microlensing
parameters are drawn uniformly in the following intervals: time of
maximum magnification $t_0$ within the observing period $\pm
300$~days, impact parameter normalized to the Einstein radius $u_0 \in
[0,1.5]$ and time-scale $t_{\rm E} \in [0,1200]$~days.
 
\begin{figure} [h] 
  \resizebox{\hsize}{!}{\includegraphics{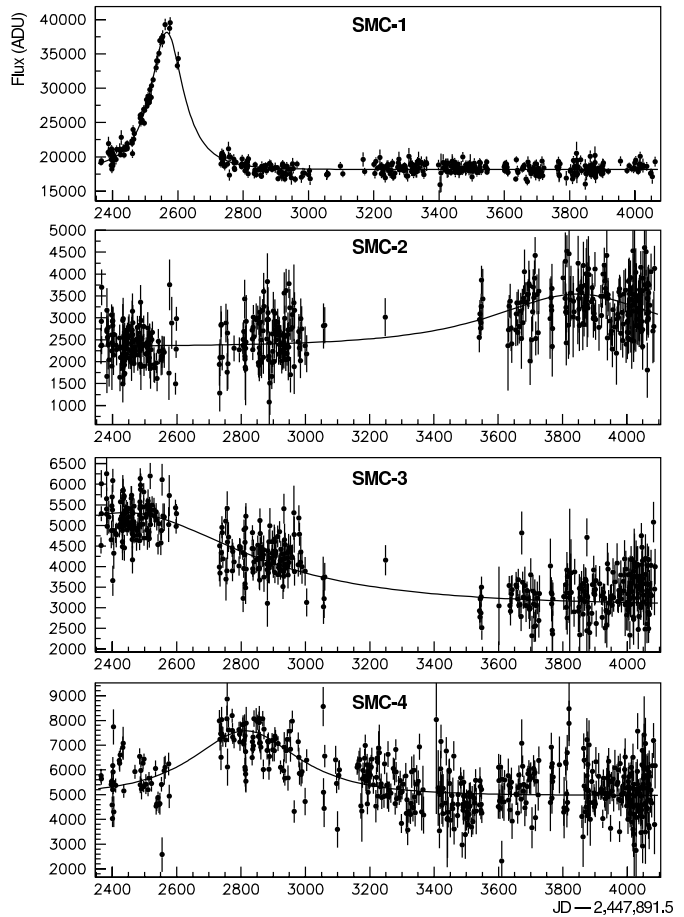}}
  \caption{Red light curves of the microlensing candidates detected
  towards the \smc.}  
  \label{fig:lcsm5} 
\end{figure}

Four candidates (\smc-1 to \smc-4) pass all the cuts. Their light
curves are shown in figure~\ref{fig:lcsm5}, their coordinates are
given in table~\ref{tab:coord} and the microlensing fit parameters (point-source,
point-lens and no-blending) in table~\ref{tab:evt}. 

\begin{table}[h]
\centering
\caption{Coordinates (J2000) and baseline photometry of the candidates.}
\label{tab:coord}
\begin{tabular}{lcccc}
\hline
   & $\alpha$ & $\delta$ & $V_J$ & $I_C$ \\
\hline
\smc-1 & 01:00:05.73 &  -72:15:02.33 & 17.9 & 17.9\\ 
\smc-2 & 00:48:20.16 &  -74:12:35.31 & 20.4 & 19.5\\ 
\smc-3 & 00:49:38.68 &  -74:14:06.77 & 20.2 & 19.3 \\ 
\smc-4 & 01:05:36.56 &  -72:15:02.39 & 19.9 & 19.4\\ 
\hline
\end{tabular}
\end{table}

\begin{table}[h]
\centering
\caption{Results of microlensing fits to the \smc\ candidates. $u_0$ is
the impact parameter and $t_{\rm E}$ the Einstein radius crossing time in
days.}\label{tab:evt}
\begin{tabular}{lccccc}
\hline
   &$u_0$ & $t_{\rm E}$ & $\chi^2/{\rm d.o.f.}$ \\
\hline
\smc-1 & 0.52 & 101 & 1026/706 \\ 
\smc-2 & 0.82 & 390 & 1526/990 \\ 
\smc-3 & 0.66 & 612 & 2316/968 \\ 
\smc-4 & 0.80 & 243 & 2923/951 \\ 
\hline
\end{tabular}
\end{table}

\begin{figure} [h] 
  \resizebox{\hsize}{!}{\includegraphics{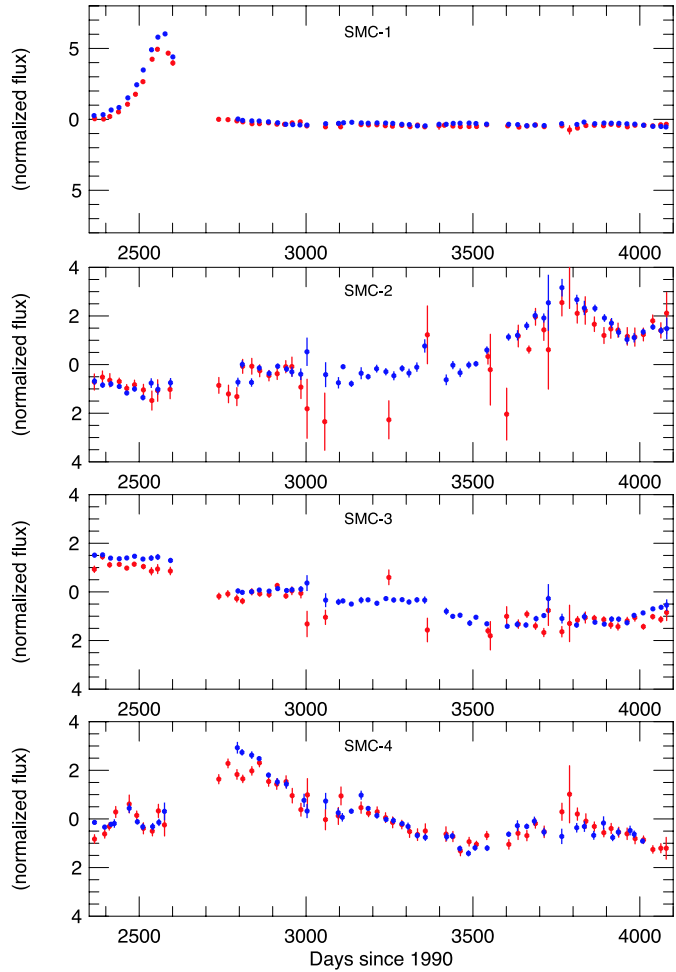}}
  \caption{Red and blue light curves of the microlensing candidates detected
  towards the \smc, in 25-day bins.}  
  \label{fig:cdl_binned} 
\end{figure}

Candidate \smc-1 is the one already present in the previous analyses
(one year and two years of \smc\ data) done by \eros. Further analysis
of its light curve indicated that it was probably a self-lensing event
(\sm98).  It was also detected as a possible microlensing
candidate by the online trigger system of the \macho\ experiment
\citep{smc1macho}. Candidates \smc-2, \smc-3 and \smc-4 have very long
durations, with few points outside the amplified region of the light curve.  The $\chi^2/{\rm d.o.f.}$ of candidate \smc-4 is large, yet far
from the cut level (which is set at 5.0 in the peak region). All three candidates, however, exhibit features in their baseline that are reminiscent of variable stars. These features appear clearly on the light curves plotted in 25-day bins (see figure~\ref{fig:cdl_binned}). To allow a direct comparison of the red and blue fluxes, all fluxes have been normalized so that they would have a mean of zero and a variance of unity. Candidates \smc-2 and \smc-4 clearly look like irregular variables, while \smc-3 has an irregular light curve and seems to be rising at the end.
These candidates will need to be monitored for several more years to be either 
confirmed but most probably ruled out as microlensing candidates. They are not in the $\sim$3
square degrees of the \smc\ on which the \macho\ project took
data.

The efficiency of the analysis for events with (Monte-Carlo) 
$u_0<1.5$ normalized to
an observing period $T_{\rm obs}$ of five years is summarized in
table~\ref{tab:eff}. 
\begin{table}[ht]
\centering
\caption{Detection efficiency $\epsilon$ in \% as a function of the
timescale $t_{\rm E}$ in days for events with $u_0<1.5$ and normalized to
$T_{\rm obs}=5{\rm \: yr}$.}
\label{tab:eff} 
{
\setlength{\tabcolsep}{4.7pt}
 \begin{tabular}{lcccccccccc} 
\hline
$t_{\rm E}$ & 5 & 15 & 50 & 125 & 325 & 500 & 900 & 1100  &1300\\ \hline
$\epsilon$ & 2.1 & 6.3 & 11.1 & 13.8 & 11.8 & 10.1 & 7.5 & 5.9 & 2.7\\ \hline
\end{tabular}
}
\end{table}

\section{Blending effect}

The above efficiencies have been derived without taking account of
stellar blending. 

We have performed a study to quantify precisely the influence of
blending on our results. We simulate an experiment with synthetic
images, regularly spaced in time, in which microlensing events are embedded. 
The images are built from the {\sc hst} colour-magnitude diagram (CMD) and 
luminosity function (LF) of the Magellanic Clouds, which extend well beyond the 
\eros\ detection limit (down to V=23.5 in the {\sc hst} data). We have checked that the
CMD and LF that we construct from these simulated images are in good 
agreement with those obtained from our data.  All stars are lensed for a 
timescale $t_{\rm E}$ of 5~images. The impact parameter $u_0$ is drawn 
uniformly between 0 and 1.5, and $t_0$ is set so that no two stars less than 
20 pixels apart are magnified simultaneously (typical \eros\ images have a
seeing of 2.1~arcsec corresponding to a gaussian with rms of
1.5~pixel).

The standard photometry chain of \eros\ is applied to these
images, leading to the detection of 5599 {\em objects} (in this
terminology an {\em object} usually encompasses several simulated {\em
stars}) with their red and blue light curves. A total of 6304
generated microlensing events were found with $u_0<1$ and an average
timescale $\langle t_{\rm E} \rangle = 3.16$ images. The effect of
blending is readily seen from the fact that more events than objects
are detected. Of these events, 60\% are due to the brightest
star in the two pixels around the object, with recovered 
$\langle t_{\rm E} \rangle = 4.00$ images and $\langle u_0\rangle = 0.52$, close to 
the generated values. 
The remaining events are due to a fainter, blended component of the
object, with recovered $\langle t_{\rm E} \rangle = 1.89$ images, well
underestimated, and $\langle u_0 \rangle = 0.64$, clearly overestimated, 
as expected from the impact of a significant blend. The
optical depth being proportional to the number of events and to their
mean $t_{\rm E}$, an estimate of the ratio $R$ of the recovered
optical depth to the generated one is:
\[R = \frac{6304}{5599/1.5}\frac{3.16}{5.0} = 1.07\;, \]
with an estimated error of about 10\%.

The light curves from the above blending simulation are then used as
templates in our Monte Carlo chain designed to compute the efficiency
of our analysis (section~3) as follows: for each object detected in the actual
data, an object is randomly chosen from the blending simulation in the
same region of the CMD. The light curve of the latter is
shifted and stretched to account for the randomly selected duration
and time of maximum amplification, but the amplitudes in each colour
are of course kept as found in the blending simulation to reflect the
impact of the underlying stellar companions of the microlensed
star. This template is used (instead of the standard shape derived by 
Paczy\'nski in the absence of  blending and used for the efficiency 
computation in the previous section) to add a microlensing event to the real
(flat in most cases) light curve. To ensure similar photometric
dispersion in the Monte Carlo and in the data, the rms flux deviation of
each simulated data point from the blended microlensing light curve is
taken to be the same as the deviation of the real light curve from
constant brightness.  The modified light curve goes through the same
analysis cuts as in the standard computation of the efficiency.

The efficiency with the blending simulation described above is within
10\% of the efficiency computed from the standard procedure, in
agreement with estimates in previous works~\citep{thn}. The loss due
to the under-estimation of the magnification of a blended event (the
observed magnification on a blended object is lower than the physical
magnification of the underlying star) is almost exactly compensated by
the fact that an object includes more than one star subject to 
lensing. In addition, as expected, the cuts of the analysis are
sufficiently loose not to reject blended microlensing events. In
particular, there are no cuts requiring a strict achromaticity of the
event. For the \eros\ experiment, blending has a small impact on
the efficiency and can therefore be neglected.

\section{Limits}

In order to set limits on the contribution of dark objects to the
halo, we use the
so-called ``standard'' halo model described in \sm98
 as model 1, and take into account the efficiency of the
analysis given in section~\ref{sec:analysis}.

For a given experiment, assuming that the halo is made of compact
objects having a single mass $M$, we define $P_{M}(t_{\rm E})$ as the
probability distribution of expected event durations, taking efficiencies into
account, and $f \tilde{N}_{\negthinspace M}$ as the expected number of
events, where $f$ is the halo mass fraction with respect to a full standard
halo. We construct a frequentist confidence level $1-p_n$ considering
only the number $n$ of observed events as a Poisson process:
\[1 - p_n = \sum_{i=0}^n\frac{e^{-f \tilde{N}_{\negthinspace M}} ( f \tilde{N}_{\negthinspace M})^i}{i!}\]

We also make an independant Kolmogorov test comparing the observed
event durations to the expected $P_{M}(t_{\rm E})$  distribution, and obtain a Kolmogorov
probability $p_t$.

A well known prescription for the confidence level of the combined test is
\[p = p_n p_t ( 1 - \ln p_n p_t)\]

Our experiment has however an unknown contribution of background --- variable stars
or any other  unknown phenomenon --- to
the observed candidates. We therefore test all combinations, attributing candidates either to signal
or background, and retain, for every mass fraction $f$, the configuration that
maximizes $p$, i.e. the most conservative one. This ensures that we have a given frequentist
coverage for all possible hypotheses.

We then set a frequentist 95\% CL limit on $f$, taking into account both the number
of candidates and their duration, by finding the value of $f$ that yields $p_{max} = 0.05$.

Although this prescription could be used to combine several experiments, it gives the
same weight to all and does not yield useful results if the sensitivities of some of the
experiments differ significantly. Therefore, to combine various experiments, we use instead a Liptak-Stouffer \citep{lip58} 
prescription where the flat p-values are converted to unit-variance, zero-mean gaussian distributed variables. A  weighted
sum of these variables is also gaussian distributed, and can be converted back to a p-value.
We take as weights the expected number
of events of each experiment for each lens mass\footnote{The final limit has a very weak dependance on the exact weighting used, provided unsensitive experiments get a very small weight.}. 

The results obtained by the various phases of \eros\ (which are independent experiments) are summarized in table~\ref{tab:evt} for microlensing candidates towards the \smc\ ---~no \smc\ candidate in \eros1~\citep{ren98}, 4 candidates in \eros2 (\sm98 and present work)~--- and in table~\ref{tab:cands} for microlensing candidates towards the \lmc\ ---~no planetary mass candidate~\citep{ren97,ren98}, \lmc-1 from \eros1 photographic plates~\citep{ans96} and the others from \eros2~\citep{las00, lasphd, mil00}.
\begin{table}
\centering
\caption{Summary of the microlensing candidates detected by \eros\ towards the \lmc. The parameters are the same as in table~\ref{tab:evt}.}
\label{tab:cands}
\begin{tabular}{lcc}
\hline
   &$u_0$ & $t_{\rm E}$ \\
\hline
\lmc-1 & 0.44 & 23 \\
\lmc-3 & 0.21 & 44 \\
\lmc-5 & 0.59 & 24 \\
\lmc-6 & 0.41 & 35 \\
\lmc-7 & 0.30 & 30 \\
\hline
\end{tabular}
\end{table}

\begin{figure} [h] 
  \resizebox{\hsize}{!}{\includegraphics{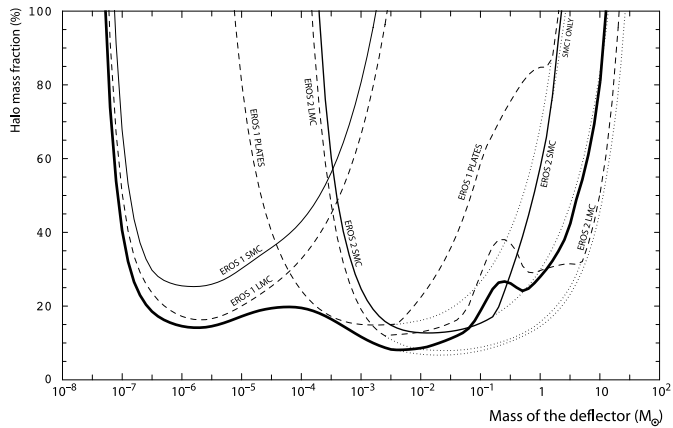}}
 \caption{ Exclusion diagram at 95~\% C.L. for the
  standard halo model ($4 \times 10^{11}\:{\rm M}_\odot$ inside
  50~kpc).  The dashed lines are the limits towards the \lmc\ by
   \eros\ 1 and \eros\ 2, the thin plain lines are limits towards the \smc\,
  and the thick line is the combined limit from the five \eros\ 
  sub-experiments. The dotted lines are the limits that would be obtained considering
 no observed events: they indicate the overall sensibility of the \eros\ dataset. We also
 indicate the limit that would be obtained if the three very long duration events on the \smc\
 were considered as background (label ``\smc-1 only'') : the limit is lowered in a negligible way in the 2--3~$\rm M_\odot$ mass range. } 
  \label{excl}
\end{figure}

Figure \ref{excl} shows the 95~\%
exclusion limit derived from this method on $f$, the halo
mass fraction, at any given mass $M$ ---~\ie\ assuming all deflectors in
the halo have mass $M$~--- for the \eros 1 \ccd\ \lmc\ and \eros 1 \ccd\
\smc, \eros 1 photographic plates, 
\eros 2 3-year \lmc\ and \eros2 5-year \smc\ experiments, and for the combination of all.
We also show in the figure the limits that would be obtained with a single \smc\ event (\smc-1), then with  no event at all in any of the \eros\ experiments, which indicates the
overall sensibility of the \eros\ project, considering presently analysed data.

The ``dent'' in the \eros1 plate limit and in the \eros2 \lmc\ limit at a mass near $0.5\rm M_\odot$ is the impact of the $\sim$30 day candidates observed towards the \lmc.
For any mass between $2 \times 10^{-7}$ and $10^{-1}~ {\rm M}_\odot$, we exclude
at 95~\% C.L. that more than 20\% of the mass of a standard halo be made of
compact objects. It can be seen that 
the combined limit is {\em above} the best limit for some values of the mass. 
This occurs quite naturally since the observed \smc\ and \lmc\ candidates have quite different characteristics. At high mass, for instance, our method will consider
\smc\ candidates as signal and \lmc\ candidates as background, thus weakening the limit obtained by the \lmc\ alone. It illustrates the
(marginally significant) incompatibility between the candidates observed by \eros\ towards the
\smc\ and the \lmc.

\section{Comparison of  {\small LMC} and {\small SMC} results}

Another way to gauge the present results on microlensing
towards the \smc\ is to compare them directly to the result
of the \macho\ collaboration towards the \lmc~\citep{lmc5macho}. 
 Such a comparison is made easier by the fact that
the expected $t_E$ distributions of halo microlenses towards
the \smc\ and \lmc\ are very similar: their averages 
and widths differ by only a few percent, and these differences
are much smaller than the widths themselves.  The ratio of the optical 
depths towards the \smc\ and \lmc\ is more uncertain; for 
a spherical isothermal halo, it is expected to be about
$\tau_{SMC} / \tau_{LMC} \simeq 1.4$, but this value can be
lower for flattened halos.  Actually, it was proposed to use 
observations towards the \smc\ to evaluate the galactic 
dark halo flattening~\citep{sacgou93}. 

In their analysis A, the \macho\ group has listed 13 microlens
candidates with an average duration $<t_E> \, = 36$~days and a width
of the $t_E$ distribution
compatible with all microlenses having the same mass.  From this,
they compute an optical depth $\tau = 1.2 \times 10^{-7}$, of which about
$1.0 \times 10^{-7}$ is attributed to halo lenses.  This translates into an
expected number of events for the present \eros\ \smc\ data set 
of 
$$ N_{evts} = N_* \frac{T_{obs}}{<t_E>} \frac{2}{\pi} <\epsilon> 
    \tau_{SMC} \, ,$$
where $N_* = 5.2 \, 10^6$ is the number of \smc\ stars monitored 
by \eros, $T_{obs} = 5$~yrs is the total duration of the observations,
and $<\epsilon> = 0.146$ is the average \eros\ detection efficiency for 
microlensing events of duration $<t_E> \simeq \, 0.1 $~yr. (This average 
efficiency is obtained by interpolating the values in Table~3, and is
normalized to $u_0 < 1$.) 
Substituting these numerical values leads to 
$N_{evts} = 2.4 \times (\tau_{\rm SMC} / 10^{-7}) $.

There are, however, no \eros\ \smc\ microlensing candidates
that are compatible with the $t_E$ distribution observed by the 
\macho\ group towards the \lmc: candidate \smc-1, 
the shortest of all, is very likely due to a lens in the \smc\
(see parallax analysis and the discussion that follows in \sm98);
the other 3 candidates 
all have durations longer than 240~days, clearly incompatible with
an average of 36~days. From this absence of microlensing candidates
with the required duration, \eros\ can exclude 
$ \tau_{SMC} > 10^{-7} $ for events similar to those in \macho\
sample A, at better than 90\%~C.L. The value 
corresponding to the spherical isothermal halo model, 
$ \tau_{SMC} = 1.4 \times \tau_{LMC} \simeq 1.4 \times10^{-7} $ is excluded at
better than 96\%~C.L. 

\section{Discussion and conclusions}

The analysis of five years of microlensing data towards the \smc\
yielded four candidates, and allowed to put stringent limits on the amount
of galactic dark matter made of compact objects. Objects with a mass
between $2\times10^{-7}\,\rm M_\odot$ and 1 $\rm M_\odot$ cannot
account for more than 25\% of the mass of a standard spherical,
isothermal and isotropic Galactic halo of $4 \times 10^{11} \,\rm
M_\odot$ out to 50~kpc.

The new method described in this paper to combine the results and use
all the information available on the events has one drawback compared
to methods used in previous works: it could be more sensitive to the
halo model, adding an extra dependence on the velocity
distribution. Previously, only the influence of the halo model on the number of
expected events has been studied, as in \sm98. 
A change in the velocity distribution, however, would
mainly shift the ``dent'' in the \eros2 \lmc\ limit. In addition, since the
final exclusion limit is quite flat over a large range of masses, the
final result should not be influenced by reasonable changes in the
velocity distribution assumptions.

This limit excludes a sizeable fraction of the allowed domain shown
in Fig.~12 of Alcock et al. (2000).

All the candidates that have been detected so far towards the \smc\ have
long durations, and seem more compatible with unidentified variable stars 
or self-lensing within the cloud than with halo objects. These would need to have supersolar masses to account for such durations. 

The statistics are still very low for a quantitative comparison of the \smc\ and \lmc\ samples of microlensing  candidates. They have, however, 
quite different characteristics which is poorly compatible with a unique population of lenses.

\begin{acknowledgements}
We thank J. Bouchez for very useful discussions.
We are grateful to D. Lacroix and the technical staff at the Observatoire de
Haute Provence and to A. Baranne for their help in refurbishing the MARLY
telescope and remounting it in La Silla. We are also grateful for the support
given to our project by the technical staff at ESO, La Silla. We thank
J.F. Lecointe for assistance with the online computing. AG was supported by grant 
AST 02-01266 from the NSF.
\end{acknowledgements}

\end{document}